\documentclass{article}

\usepackage{arxiv}

\usepackage[utf8]{inputenc} 
\usepackage[T1]{fontenc}    
\usepackage{url}            
\usepackage{booktabs}       
\usepackage{amsfonts}       
\usepackage{nicefrac}       
\usepackage{microtype}      
\usepackage{graphicx}
\usepackage[numbers]{natbib}
\usepackage{doi}

\usepackage{comment}
\usepackage{amsmath,amssymb,amsfonts,amsthm}
\usepackage{dirtytalk}
\usepackage{titling}
\renewcommand{\v}{\mathbf}

\newcommand{\bm}{\begin{bmatrix}}
\newcommand{\ema}{\end{bmatrix}}

\usepackage{hyperref}       

\title{A direct approach to solving the hydrodynamic oseenlet}

\date{November, 2021}	

\thanksmarkseries{alph}
\author{ Aditya R. Pujari \thanks{Alternate email: adi57pujari@gmail.com} \\ 
\small	International Centre for Theoretical Sciences - TIFR\\ \small Shivakote, Bengaluru (North) - 560 089, India. \\ \small \texttt{aditya.pujari@icts.res.in} \\
}

\renewcommand{\headeright}{}
\renewcommand{\undertitle}{}
\renewcommand{\shorttitle}{A direct approach to solving the hydrodynamic oseenlet}

\hypersetup{
pdftitle={A direct approach to solving the hydrodynamic oseenlet},
pdfsubject={q-bio.NC, q-bio.QM},
pdfauthor={Aditya R. Pujari},
pdfkeywords={Stokes flow, Oseen flow, Stokeslet, Oseenlet, Point Force Singularities},
}

\begin{document}
\maketitle

\begin{abstract}
	 We present a new direct approach to obtain the Green's function of the Oseen equations, also known as the oseenlet. This approach is different in that it does not assume an initial form of the solution to begin with. The final solution is expressed in a new, simplified form that resembles the Oseen-Burger's tensor of the stokeslet equations, from which the contributions due to inertia are easily appreciable.
	
\end{abstract}


\section{Introduction}

Green's functions are useful mathematical constructs, especially in fluid dynamics while describing low Reynolds number flows, such as Stokes flow and Oseen flow. Stokes flow in particular has been extensively studied in literature over the past several decades. Various methods have been devised to obtain the Green's function of Stokes flow\textemdash also known as Oseen-Burger's tensor, or the stokeslet \cite{lisicki2013four}. Oseen flow, however, is a relatively new and expanding subject, and literature on obtaining the Green's function of Oseen flow\textemdash such as that given by Pozrikidis \cite{pozrikidis2011introduction}\textemdash is limited. 

Here we present a new direct approach to obtain the Green's function of the Oseen equations, also popularly known as the oseenlet, using Fourier transformations. This approach differs from that of Pozrikidis, in that it does not assume a certain form of the solution to begin with. The motivation to come up with a solution of the oseenlet equations using Fourier transformations came from appreciating the elegance of the method of Zapryanov and Tabakova to solve the stokeslet \cite{zapryanov2013dynamics} (hereafter, referred to as the Z\&T method). Their method makes use of the Fourier transforms of the fundamental solutions of known partial differential equations; as a result, it requires little calculation and the solution is obtained within a page of simple algebra. Intrigued by their method, a similar direct approach was devised to solve the oseenlet. 

\section{The Oseen Equations and the Oseenlet}
The Oseen equations are a linearized approximation of the Navier-Stokes equations. We briefly derive them here for the steady case. Let $\v{u}$ and $p$ denote the velocity and pressure of the fluid in the domain, which are functions of the spatial variable $\v{x}$. The Navier-Stokes equations are given by,
\begin{equation}
    \rho\v{u}(\v{x})\cdot\v{\nabla}\v{u}(\v{x}) = -\v{\nabla}p(\v{x}) + \mu \nabla^2 \v{u}(\v{x}) 
\end{equation}
\begin{equation}
\label{eqn:NSdiv}
    \v{\nabla}\cdot\v{u}(\v{x}) = 0
\end{equation}
where $\rho$ is the density of the fluid and $\mu$ is the dynamic viscosity of the fluid. Let us decompose $\v{u}$ into a constant average flow $\v{U}$ and a perturbed flow $\v{v}(\v{x})$.
\begin{equation}
    \v{u} = \v{U} + \v{v}(\v{x})
\end{equation}
Rewriting the Navier-Stokes equations in terms of this decomposition we get,
\begin{equation}
    \rho(\v{U} + \v{v}(\v{x}))\cdot\v{\nabla}(\v{U} + \v{v}(\v{x})) = -\v{\nabla}p(\v{x}) + \mu \nabla^2 (\v{U} + \v{v}(\v{x}))
\end{equation}
\begin{equation}
    \nabla \cdot (\v{U} + \v{v}(\v{x})) = 0.
\end{equation}
On expanding the terms, we see that the term $\rho \v{v}\cdot\nabla\v{v}$ may be neglected as a higher order $O(v^2)$ term. This gives the Oseen approximation. 
\begin{equation}
    \rho\v{U}\cdot\nabla\v{U} + \rho\v{U}\cdot\nabla\v{v}(\v{x}) + \rho\v{v}(\v{x})\cdot\nabla\v{U} + \rho\v{v}(\v{x})\cdot\nabla\v{v}(\v{x}) = -\nabla p + \mu\nabla^2\v{U} + \mu\nabla^2\v{v} +\v{b}\delta(\v{x}) 
\end{equation}

In the present case where we wish to describe the flow due to a point force singularity, we obtain the equations of the oseenlet by adding the point force term, given by,

\begin{equation}
\label{eqn:oseenlet}
    \rho\v{U}\cdot\v{\nabla}\v{v}(\v{x}) = -\v{\nabla}p(\v{x}) + \mu \nabla^2 \v{v}(\v{x}) + \v{b}\delta(\v{x})
\end{equation}
\begin{equation}
\label{eqn:divergence}
    \v{\nabla}\cdot\v{v}(\v{x}) = 0.
\end{equation}
Here $\v{b}$ is the point force located at the origin and $\delta(\v{x})$ is the 3-dimensional delta function. The boundary conditions are that $\v{v}(\v{x}) \rightarrow 0$ and $p(\v{x}) \rightarrow 0$ as $\v{x} \rightarrow \infty$. 

\section{Solution}
To solve the equations of the oseenlet, we begin, as in the Z\&T method, by taking the Fourier transform of eq. \ref{eqn:oseenlet} and eq. \ref{eqn:divergence}. Refer to the appendix for the definition of the Fourier transform.  
\begin{equation}
\label{eqn:oseenlet k space}
    i\rho \v{U} \cdot \v{k} \, \v{\hat{v}} = -i\v{k}\hat{p} - \mu k^2 \v{\hat{v}} + \v{b} 
\end{equation}
\begin{equation}
\label{eqn:divergence k space}
    i\v{k}\cdot\v{\hat{v}} = 0 
\end{equation}
Taking the dot product of eq. \ref{eqn:oseenlet k space} with $\v{k}$, and using eq. \ref{eqn:divergence k space} to eliminate the two velocity terms, we rearrange to obtain the pressure solution in k-space. This is equivalent to taking the divergence of eq. \ref{eqn:oseenlet} and eliminating the velocity terms due to the continuity equation (eq. \ref{eqn:divergence}). We are able to do this trick owing to the linearity of the equation in $\v{v}$; had the inertial term not been linearized, it would not have been eliminated on taking the divergence. 
\begin{equation}
\label{eqn:pressure k space}
    \hat{p} = -i \frac{\v{k}\cdot\v{b}}{k^2}
\end{equation}
On inverting this, we obtain,
\begin{equation}
    p = -\frac{1}{4\pi}\nabla\left(\frac{1}{r} \right) \cdot \v{b} = - \frac{1}{4\pi} \frac{\v{x}}{r}\cdot \v{b}.
\end{equation}
Here $r=|\v{x}|$. The pressure solution is the same as that of the stokeslet, as also pointed out by Pozrikidis. Re-substituting the pressure solution into eq. \ref{eqn:oseenlet k space} we obtain an expression for velocity in k-space. 
\begin{eqnarray}
    i\rho \v{U}\cdot\v{k} \, \v{\hat{v}} = - \frac{\v{k}\v{k}\cdot\v{b}}{k^2} - \mu k^2 \v{\hat{v}} + \v{b} \\
\label{eqn:velocity k space}
    \v{\hat{v}} = \frac{1}{\frac{i\v{U}\cdot\v{k}}{\nu} + k^2} \left[ \frac{\v{b}}{\mu} - \frac{\v{k}\v{k}\cdot\v{b}}{\mu k^2} \right]
\end{eqnarray}
Here $\nu$ is the kinematic viscosity of the fluid. We see that in the k-space velocity solution given by eq.  \ref{eqn:velocity k space}, if $\v{U} \rightarrow 0$, we obtain the corresponding k-space velocity solution of the stokeslet equations $\frac{\v{b}}{\mu k^2} + \frac{\v{k}\v{k}\cdot\v{b}}{\mu k^4}$. Note that the stokeslet solution is symmetric in $\v{k}$, however, in the oseenlet the $\frac{i\v{U}\cdot\v{k}}{\nu}$ term introduces an asymmetry in the velocity solution. We complete the square in the denominator and see that, $k^2 + \frac{i\v{U}\cdot\v{k}}{\nu} = \left( \v{k} + \frac{i\v{U}}{2\nu} \right)^2 + \left(\frac{|\v{U}|}{2\nu}\right)^2$. The completed square may be treated as a shift in k-space by the constant $\frac{i\v{U}}{2\nu}$. Thus, to invert the first term, we seek to find a scalar function in this shifted k-space whose Fourier transform is the fraction $\frac{1}{k^2+c^2}$, where $c$ is some constant. This leads us to the Green's function of the Helmholtz equation, given by:
\begin{equation}
    \nabla^2\Phi(\v{x}) - c^2\Phi(\v{x}) + \delta(\v{x}) = 0.
\end{equation}
Thus we have, 
\begin{equation}
    \hat{\Phi}(\v{k}) = \frac{1}{k^2+c^2}
\end{equation}
and, 
\begin{equation}
    \hat{\Phi}(\v{k}+\tfrac{i\v{U}}{2\nu}) = \frac{1}{k^2 + \frac{i\v{U}\cdot\v{k}}{\nu}}.
\end{equation}
Here, $c$ is $\frac{|\v{U}|}{2\nu}$. We know that the solution of the Helmholtz equation is 
\begin{equation}
    \mathcal{F}^{-1}\left\{ \hat{\Phi}(\v{k}) \right\} = \frac{1}{4\pi r}e^{-cr}
\end{equation}
from which, a shift in k-space corresponds to an exponential factor appearing in x-space,
\begin{equation}
    \mathcal{F}^{-1}\left\{ \hat{\Phi}(\v{k} + \tfrac{i\v{U}}{2\nu}) \right\} = \frac{1}{4\pi r} e^{-\frac{|\v{U}|r}{2\nu}}e^{\frac{\v{U}\cdot\v{x}}{2\nu}}.
\end{equation}
Thus we have the inversion of the first term. 
\begin{equation}
    \v{v}_1 = \frac{\v{b}}{4\pi\mu r} e^{\frac{\v{U}\cdot\v{x}}{2\nu}-\frac{|\v{U}|r}{2\nu}}
\end{equation}
Before we move on to inverting the second term, it is notable that we have naturally been graced by a dimensionless variable in the form of the exponent. We choose to call this dimensionless variable as $\eta$ and expect to encounter it again later while solving the second term. ($\eta$ is defined with a negative sign to align with Pozrikidis' notation.)
\begin{equation}
\label{eqn:eta}
    \eta \,\,\dot{=}\,\, \frac{|\v{U}|r}{2\nu} - \frac{\v{U}\cdot\v{x}}{2\nu} 
\end{equation}
The second term in the velocity expression is not as straightforward as the first. Let us first rewrite the second term in terms of $\hat{\Phi}$. 
\begin{equation}
    \v{v}_2 = - \hat{\Phi}(\v{k}+\tfrac{i\v{U}}{2\nu}) \frac{\v{k}\v{k}\cdot\v{b}}{\mu k^2}
\end{equation}
If, somehow, we are able to obtain the inversion of the function $\hat{\Phi}(\v{k}+\tfrac{i\v{U}}{2\nu})/k^2$, then we would only be left with calculating a dyadic differentiation after inversion. Let us call this function $\hat{H}$. Thus, 
\begin{eqnarray}
\label{eqn:h hat}
    \hat{H} \,\,\dot{=}\,\, -\frac{\hat{\Phi}(\v{k}+\tfrac{i\v{U}}{2\nu})}{k^2} \\
    \v{\hat{v}}_2 = \frac{\v{k}\v{k}\hat{H}\cdot\v{b}}{\mu} \\
    \v{v}_2 = -\frac{\v{\nabla}\v{\nabla}H\cdot\v{b}}{\mu}. \\
\end{eqnarray}
We simplify for $\hat{H}$,
\begin{eqnarray}
    k^2 \hat{H} = -\hat{\Phi}(\v{k}+\tfrac{i\v{U}}{2\nu}) \\
\label{eqn:Laplacian H 1}
    \nabla^2 H = \frac{1}{4\pi r} e^{-\eta} .
\end{eqnarray}
Since it is not possible to integrate $\nabla^2 H$ directly, we work backwards starting from $H$ instead. We anticipate that $H$ will be a function of the dimensionless variable $\eta$ obtained above. With this assumption, we proceed to differentiate $H$. From here, the solution follows in the footsteps of Pozrikidis' solution. 
\begin{equation*}
    H = H(\eta)
\end{equation*}
\begin{equation}
\label{eqn:gradient H}
    \v{\nabla} H = \frac{|\v{U}|}{2\nu} \left( \frac{\v{x}}{r} - \v{e} \right) \frac{dH}{d\eta} 
\end{equation}
Here $\v{e}$ is the unit vector in the direction of $\v{U}$. Calculating the Laplacian of $H$:
\begin{eqnarray}
\label{eqn:Laplacian H 2}
    \nabla^2 H \equiv \v{\nabla}\cdot\v{\nabla}H = \frac{|\v{U}|}{2\nu} \frac{d}{d\v{x}} \left[ \frac{dH}{d\eta} \left( \frac{\v{x}}{r} - \v{e} \right)  \right] = \frac{|\v{U}|}{\nu} \frac{1}{r} \left( \eta \frac{d^2 H}{d\eta^2} + \frac{dH}{d\eta}\right) .
\end{eqnarray}
Equating the expression for $\nabla^2 H$ in eq.  \ref{eqn:Laplacian H 2} with that in eq. \ref{eqn:Laplacian H 1} and integrating we get,
\begin{align}
    \frac{1}{4\pi r}e^{-\eta} & = \frac{|U|}{\nu}\frac{1}{r}\frac{\partial}{\partial\eta}\left(\eta\frac{\partial H}{\partial\eta}\right) \\
    -\frac{1}{4\pi} \left.\frac{\nu}{|U|}e^{-\eta} \right|^{\eta}_{0} & = \eta\frac{\partial H}{\partial\eta} \;\;\;\;\;\;\;  \text{(since $\eta$ is always positive)}\\
    H & = \frac{1}{4\pi} \frac{\nu}{|U|} \int^{\eta}_{0} \frac{1-e^{-\eta'}}{\eta'} d\eta'.
\end{align}
Thus we have successfully obtained the function $H$. What remains is only the dyadic differentiation of $H$ in order to calculate the second term. 
\begin{align}
    \v{v}_2 & = -\frac{\v{b}\cdot\v{\nabla}}{\mu} \v{\nabla}H \\
    & = -\frac{\v{b}\cdot\v{\nabla}}{\mu} \left[ \frac{\partial H}{\partial\eta}\left( \frac{|\v{U}|}{2\nu} \left( \frac{\v{x}}{r} - \v{e} \right) \right) \right] \\
    \v{v}_2 & = -\frac{\v{b}\cdot\v{\nabla}}{8\pi\mu} \left[ \left( \frac{1-e^{-\eta}}{\eta} \right)\left( \frac{\v{x}}{r} - \v{e} \right) \right] 
\end{align}
Finally, we put together the two terms to obtain the oseenlet velocity solution, which is in agreement with the solution obtained by Pozrikidis. 
\begin{equation}
    \v{v}(\v{x}) = \frac{1}{8\pi\mu} \left\{ \frac{2\v{b}}{r}e^{-\eta} -\v{b}\cdot\v{\nabla} \left[ \left( \frac{1-e^{-\eta}}{\eta} \right)\left( \frac{\v{x}}{r} - \v{e} \right) \right]  \right\}.
\end{equation}

\section{Discussion}

We continue to express the oseenlet solution in a more tractable format. On carrying out the differentiation of the second term, we obtain:
\begin{equation}
    \v{u}(\v{x}) = \frac{1}{8\pi\mu} \mathbb{O}\cdot\v{b}
\end{equation}
where, 
\begin{equation}
    \mathbb{O} \,\, \dot{=} \,\, \frac{2}{r}e^{-\eta}\mathbb{I} - \frac{U}{2\nu}\left(\frac{ \eta e^{-\eta} + e^{-\eta} - 1 }{\eta^2}\right) \left( \frac{\v{x}}{r} - \frac{\v{U}}{U} \right) \left( \frac{\v{x}}{r} - \frac{\v{U}}{U} \right) + \left( \frac{e^{-\eta} - 1}{\eta} \right) \left( \frac{\mathbb{I}}{r} - \frac{\v{x}\v{x}}{r^3} \right).
\end{equation}
Here $\mathbb{I}$ is the identity tensor. The tensor may be rewritten in a manner so as to resemble the Oseen-Burger's tensor of the stokeslet:
\begin{equation}
    \mathbb{O} = \eta_1 \frac{\mathbb{I}}{r} + \eta_2 \frac{\v{x}\v{x}}{r^3} + \frac{U}{2\nu}\eta_3 \v{x}^\dagger \v{x}^\dagger 
\end{equation}
where the dimensionless weights $\eta_1$, $\eta_2$, $\eta_3$ and the dimensionless unit vector $\v{x}^\dagger$ are defined as, 
\begin{align}
    &\eta_1 \,\, \dot{=} \,\, \left( \frac{2\eta e^{-\eta} + e^{-\eta} - 1 }{\eta}\right) 
    &\eta_2 \,\, \dot{=} \,\, \left( \frac{1 - e^{-\eta}}{\eta}\right) &
    &\eta_3 \,\, \dot{=} \,\, \left( \frac{1 - \eta e^{-\eta} - e^{-\eta}}{\eta^2} \right) \\
    &\v{x}^\dagger \,\, \dot{=} \,\, \left(\frac{\v{x}}{r} - \frac{\v{U}}{U}\right).
\end{align}
$\eta$, acts as an \emph{angular} Reynolds number, \textit{i.e.}, the Reynolds number weighted by a factor of $(1-\cos{\theta})/2$. This can be seen by manipulating the definition of $\eta$, eq. \ref{eqn:eta}:
\begin{equation}
    \eta \,\,\dot{=}\,\, \frac{|\v{U}|r}{2\nu} - \frac{\v{U}\cdot\v{x}}{2\nu} = \frac{|\v{U}|r}{2\nu}(1-\cos{\theta}) = \dfrac{1}{2}Re_r(1-\cos{\theta}).
\end{equation}
Here $\theta$ is the angle made by the position vector with the mean flow velocity direction $\v{U}$ and the length scale for the calculation of the Reynolds number $Re_r$ is taken to be magnitude of the position vector $r$. $\eta_i$, $i=1,2,3$, is thus an \emph{angular} coefficient that weights each of the terms. For a given Reynolds number $Re_r$, each $\eta_i$ is a monotonically decreasing function of $\theta$. Similarly, $\v{x}^\dagger$ is the \emph{skew} vector describing the extent to which the position vector is skewed from the mean flow velocity direction $\v{U}$. 

The first two terms are similar to the two terms of the Oseen-Burger's tensor of the stokeslet solution, and differ only by the scalar weights. As $U \rightarrow 0$, the scalar factors $\eta_1,\eta_2 \rightarrow 1$ and $\eta_3 \rightarrow 0$, appropriately recovering the stokeslet solution. 

\section{Conclusion}

In this article, we discussed a new direct approach to obtain the Green's function of the oseen equations. Without beginning with an assumption of a specific form of the solution, we made the use of Fourier transformation to first obtain the solution in $k$-space. The denominator of the first term naturally suggests using the known solution of the Helmholtz equation for inversion. A non-dimensional parameter $\eta$, which behaves as an angular Reynolds number, is naturally obtained in the exponent. To invert the second term, we work backwards by performing an integration on the lines of Pozrikidis' solution. 

The final solution is expressed in a new tractable format, similar to that of the stokeslet. The presence of the $\eta_i$ factors is a direct result of the inertia of the fluid. These $\eta_i$ factors along with the skew vector $\v{x}^\dagger$ are in turn responsible for the creation of fore-aft asymmetry in the velocity field. 

\section*{Acknowledgements}
The author acknowledges support of the Department of Atomic Energy, Government of India, under project no. RTI4001. The author is grateful to Prof. Rama Govindarajan for guidance and would also like to thank Anup Kumar and Divya Jagannathan for fruitful discussions.

\section*{Appendix A. Fourier Transforms}

We define the Fourier transform as: 
\begin{equation}\tag{A1}
\label{a:A1}
    \mathcal{F}\{{f(\v{x})}\} \,\,\dot{=}\,\, \hat{f}(\v{k}) \,\,\dot{=}\,\, \int_{\mathbb{R}^3} d\v{x} f(\v{x}) e^{-i\v{k}\cdot\v{x}}
\end{equation}
and the inverse as
\begin{equation}\tag{A2}
\label{a:A2}
    \mathcal{F}^{-1} \{{f(\v{k})}\} \,\,\dot{=}\,\, f(\v{x}) \,\,\dot{=}\,\, \frac{1}{(2\pi)^3} \int_{\mathbb{R}^3} d\v{k} \hat{f}(\v{k}) e^{i\v{k}\cdot\v{x}}.
\end{equation}
We have made use of the Fourier transform of the Green's function of the Laplace equation while calculating the pressure solution.
\begin{equation}\tag{A3}
    \nabla^2 \phi(\v{x}) = -\delta(x)
\end{equation}
where $\phi = \frac{1}{4\pi r}$. By integrating its solution, we can obtain the corresponding Fourier space expression,
\begin{align*}
\tag{A4}
    \nabla^2 \left( \frac{1}{4\pi r} \right) & = -\delta(\v{x}) = -\frac{1}{(2\pi^3)} \int_{\mathbb{R}^3} d\v{k}  e^{i\v{k}\cdot\v{x}} \\
\tag{A5}
    \nabla \left( \frac{1}{4\pi r} \right) & = -\frac{1}{(2\pi^3)} \int_{\mathbb{R}^3} d\v{k} \frac{i\v{k}}{k^2} e^{i\v{k}\cdot\v{x}} \\
\tag{A6}
    \left( \frac{1}{4\pi r} \right) & = \frac{1}{(2\pi^3)} \int_{\mathbb{R}^3} d\v{k} \frac{1}{k^2} e^{i\v{k}\cdot\v{x}} \\
\end{align*}

\bibliographystyle{abbrv} 

\begin{thebibliography}{1}

\bibitem{lisicki2013four}
M.~Lisicki.
\newblock Four approaches to hydrodynamic green's functions--the oseen tensors,
  2013.
\newblock Preprint at \url{https://arxiv.org/abs/1312.6231}.

\bibitem{pozrikidis2011introduction}
C.~Pozrikidis.
\newblock {\em Introduction to theoretical and computational fluid dynamics}.
\newblock Oxford university press, 2011.

\bibitem{zapryanov2013dynamics}
Z.~Zapryanov and S.~Tabakova.
\newblock {\em Dynamics of bubbles, drops and rigid particles}, volume~50.
\newblock Springer Science \& Business Media, 2013.

\end{thebibliography}


\relax 
\providecommand\hyper@newdestlabel[2]{}
\providecommand\HyperFirstAtBeginDocument{\AtBeginDocument}
\HyperFirstAtBeginDocument{\ifx\hyper@anchor\@undefined
\global\let\oldcontentsline\contentsline
\gdef\contentsline#1#2#3#4{\oldcontentsline{#1}{#2}{#3}}
\global\let\oldnewlabel\newlabel
\gdef\newlabel#1#2{\newlabelxx{#1}#2}
\gdef\newlabelxx#1#2#3#4#5#6{\oldnewlabel{#1}{{#2}{#3}}}
\AtEndDocument{\ifx\hyper@anchor\@undefined
\let\contentsline\oldcontentsline
\let\newlabel\oldnewlabel
\fi}
\fi}
\global\let\hyper@last\relax 
\gdef\HyperFirstAtBeginDocument#1{#1}
\providecommand\HyField@AuxAddToFields[1]{}
\providecommand\HyField@AuxAddToCoFields[2]{}
\citation{lisicki2013four}
\citation{pozrikidis2011introduction}
\citation{zapryanov2013dynamics}
\@writefile{toc}{\contentsline {section}{\numberline {1}Introduction}{1}{section.1}\protected@file@percent }
\@writefile{toc}{\contentsline {section}{\numberline {2}The Oseen Equations and the Oseenlet}{1}{section.2}\protected@file@percent }
\newlabel{eqn:NSdiv}{{2}{1}{The Oseen Equations and the Oseenlet}{equation.2.2}{}}
\newlabel{eqn:oseenlet}{{7}{2}{The Oseen Equations and the Oseenlet}{equation.2.7}{}}
\newlabel{eqn:divergence}{{8}{2}{The Oseen Equations and the Oseenlet}{equation.2.8}{}}
\@writefile{toc}{\contentsline {section}{\numberline {3}Solution}{2}{section.3}\protected@file@percent }
\newlabel{eqn:oseenlet k space}{{9}{2}{Solution}{equation.3.9}{}}
\newlabel{eqn:divergence k space}{{10}{2}{Solution}{equation.3.10}{}}
\newlabel{eqn:pressure k space}{{11}{2}{Solution}{equation.3.11}{}}
\newlabel{eqn:velocity k space}{{14}{2}{Solution}{equation.3.13}{}}
\newlabel{eqn:eta}{{21}{3}{Solution}{equation.3.21}{}}
\newlabel{eqn:h hat}{{23}{3}{Solution}{equation.3.23}{}}
\newlabel{eqn:Laplacian H 1}{{28}{3}{Solution}{equation.3.27}{}}
\newlabel{eqn:gradient H}{{29}{3}{Solution}{equation.3.29}{}}
\newlabel{eqn:Laplacian H 2}{{30}{3}{Solution}{equation.3.30}{}}
\@writefile{toc}{\contentsline {section}{\numberline {4}Discussion}{4}{section.4}\protected@file@percent }
\bibstyle{abbrv}
\bibcite{lisicki2013four}{{1}{}{{}}{{}}}
\bibcite{pozrikidis2011introduction}{{2}{}{{}}{{}}}
\bibcite{zapryanov2013dynamics}{{3}{}{{}}{{}}}
\providecommand\NAT@force@numbers{}\NAT@force@numbers
\@writefile{toc}{\contentsline {section}{\numberline {5}Conclusion}{5}{section.5}\protected@file@percent }
\newlabel{a:A1}{{{A1}}{5}{Appendix A. Fourier Transforms}{equation.5.44}{}}
\newlabel{a:A2}{{{A2}}{5}{Appendix A. Fourier Transforms}{equation.5.44}{}}
\gdef \@abspage@last{5}

\end{document}